\documentclass[nofootinbib,preprint,floatfix, longbibliography, superscriptaddress]{revtex4-1}
\usepackage{amsmath}
\usepackage{amssymb}
\usepackage{graphicx}
\usepackage{epstopdf}
\usepackage{verbatim}
\usepackage{color}
\usepackage{subfigure}
\usepackage[pdftex, breaklinks,colorlinks, citecolor=blue, urlcolor=blue]{hyperref}

\begin{document}
\title{Laser cooling of a micromechanical membrane to the quantum backaction limit}

\author{R.W.~Peterson}
\affiliation{JILA, University of Colorado and NIST, Boulder, Colorado 80309, USA}
\affiliation{Department of Physics, University of Colorado, Boulder, Colorado 80309, USA}
\author{T.P.~Purdy}
\affiliation{JILA, University of Colorado and NIST, Boulder, Colorado 80309, USA}
\affiliation{Department of Physics, University of Colorado, Boulder, Colorado 80309, USA}
\author{N.S.~Kampel}
\affiliation{JILA, University of Colorado and NIST, Boulder, Colorado 80309, USA}
\affiliation{Department of Physics, University of Colorado, Boulder, Colorado 80309, USA}
\author{R.W.~Andrews}
\affiliation{JILA, University of Colorado and NIST, Boulder, Colorado 80309, USA}
\affiliation{Department of Physics, University of Colorado, Boulder, Colorado 80309, USA}
\author{P.-L.~Yu}
\affiliation{JILA, University of Colorado and NIST, Boulder, Colorado 80309, USA}
\affiliation{Department of Physics, University of Colorado, Boulder, Colorado 80309, USA}
\author{K.W.~Lehnert}
\affiliation{JILA, University of Colorado and NIST, Boulder, Colorado 80309, USA}
\affiliation{Department of Physics, University of Colorado, Boulder, Colorado 80309, USA}
\affiliation{National Institute of Standards and Technology (NIST), Boulder, Colorado, 80305, USA}
\author{C.A.~Regal}
\affiliation{JILA, University of Colorado and NIST, Boulder, Colorado 80309, USA}
\affiliation{Department of Physics, University of Colorado, Boulder, Colorado 80309, USA}

\begin{abstract}
The radiation pressure of light can act to damp and cool the vibrational motion of a mechanical resonator.  In understanding the quantum limits of this cooling, one must consider the effect of shot noise fluctuations on the final thermal occupation.  In optomechanical sideband cooling in a cavity, the finite Stokes Raman scattering defined by the cavity linewidth combined with shot noise fluctuations dictates a quantum backaction limit, analogous to the Doppler limit of atomic laser cooling.  In our work we sideband cool to the quantum backaction limit by using a micromechanical membrane precooled in a dilution refrigerator.  Monitoring the optical sidebands allows us to directly observe the mechanical object come to thermal equilibrium with the optical bath.
\end{abstract}

\maketitle

Laser cooling revolutionized atomic physics and paved the way to the creation of extreme states of matter with ensembles of atoms. Recent achievements in optomechanics have paralleled the early development of atomic laser cooling~\cite{chan2011,teufel2011}, and raise the promise of using micromechanical resonators in a variety of quantum devices. The prospects for sensitive manipulation of macroscopic objects with light were first studied in the context of gravitational wave detection~\cite{braginsky1967,braginsky1970}. It was recognized that the radiation pressure force of intense laser light could act as a source of dissipation for a harmonically bound object, an effect termed dynamical backaction. Not until later was the quantum nature of dynamical backaction considered: Shot noise imparts a random backaction force on the resonator, termed radiation pressure shot noise (RPSN)~\cite{caves1980}. In the context of optomechanical sideband cooling, the relative rates of Stokes and anti-Stokes Raman scattering define a quantum backaction limit below which the resonator cannot be cooled~\cite{marquardt2007,wilson-rae2007}. This limit is analogous to the Doppler limit in atomic laser cooling, which is the minimum achievable temperature on a given linewidth atomic transition, due to the randomly-oriented momentum kicks from the spontaneously emitted photons~\cite{chu1985}. 

Recent work in optomechanics has explored the interaction of light and mechanical motion in the quantum regime, from observing the backaction induced by position measurement~\cite{purdy2013}, to using backaction to generate squeezed states of light~\cite{safavi-naeini2013,purdy2013a}. Although sideband cooling has allowed preparation of a mechanical resonator into its quantum ground state~\cite{chan2011,teufel2011}, a regime where the quantum nature of backaction is relevant~\cite{safavi-naeni2011b,khalili2012, palomaki2013b}, these experiments still operate far from the quantum limit of sideband cooling. In this Letter, by precooling a micromechanical membrane in a dilution refrigerator, we are able to directly observe the mechanical resonator come to thermal equilibrium with an optical bath, which for the sideband resolving power of our cavity reaches a mechanical phonon occupation of ${\bar{n}=0.20\pm0.02}$. Even at the low phonon occupation that defines the quantum backaction limit in our device, we observe no evidence of heating of the mechanical material due to optical absorption.  This is a crucial realization for future application of cryogenic optomechanical devices such as transduction between microwave and optical photons~\cite{cleland2013,andrews2014}.

\begin{figure}
\scalebox{0.7}{\includegraphics{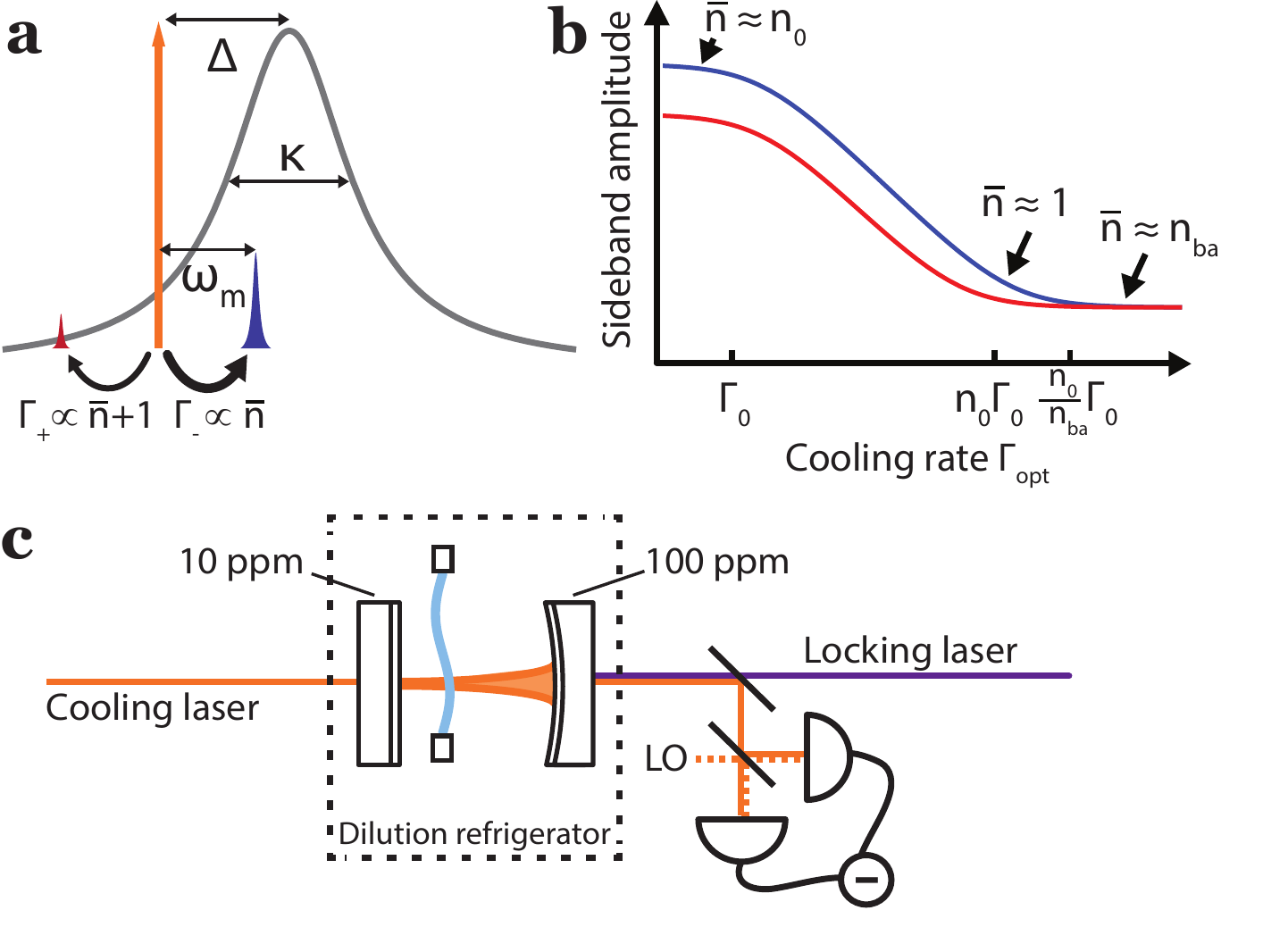}}
\caption{Optomechanical sideband cooling and Raman-ratio thermometry. a) Stokes and anti-Stokes scattering rates $\Gamma_{+}$ and $\Gamma_{-}$ depend on both the mechanical mode's phonon occupation $\bar{n}$, as well as the factors proportional to cavity susceptibility at $\Delta \mp \omega_m$ (gray curve; full linewidth $\kappa$). b) Fractional sideband amplitude for the case of backaction limit $n_{\text{ba}}<1$ (log-log scale). When negligible optical cooling is applied ($\Gamma_{\text{opt}} < \Gamma_0$) (left), the Stokes (red) and anti-Stokes (blue) sidebands correspond to the ratio of cavity susceptibility, but the cooling laser is too weak to lower the temperature. In the classical cooling regime (center), temperature decreases while the ratio of Stokes and anti-Stokes sidebands remains constant. When $\Gamma_{\text{opt}} \approx n_0 \Gamma_0$, the mechanical decoherence rate, $\bar{n}$ is approaching the ground state. At $\Gamma_{\text{opt}} \approx \frac{n_0}{n_{\text{ba}}}\Gamma_0$, backaction and thermal motion equally contribute to mechanical motion. Beyond this is the backaction limit regime (right). c) Experimental setup. The cooling laser (orange) is injected into the optical cavity through the 10 part per million (ppm) transmission mirror. Transmitted cooling laser light passes through the 100 ppm mirror and is collected in heterodyne detection with a local oscillator (LO). The cavity is actively stabilized using a locking laser (purple) injected into the orthogonal polarization mode of the cavity.} 

\label{fig:setup}
\end{figure}

In an optomechanical sideband cooling experiment, the cooling laser is red-detuned from a resonant mode of an optical cavity coupled to mechanical motion. The cavity's susceptibility enhances the near-resonant anti-Stokes scattered light, which removes mechanical quanta from the mechanical resonator when it exits the cavity, and suppresses the far-off-resonant Stokes scattered light, which adds mechanical quanta (Fig.~\ref{fig:setup}a). This net cooling effect can be strengthened by increasing the cooling laser power, with the mechanical mode's final temperature a balance between its intrinsic thermal bath temperature and the cold optical bath provided by the cooling laser and cavity.  In the so-called resolved-sideband regime, near-complete suppression of Stokes scattering~\cite{chan2011,teufel2011} can be achieved by setting the cavity linewidth $\kappa$ to be much smaller than the mechanical frequency $\omega_m$. In this limit, optical cooling can only remove energy from the mechanical system. However, for any finite sideband resolution, the cooling due to the imbalance of cavity susceptibility at each sideband is eventually undone by the fundamental asymmetry of Raman scattering. Namely, anti-Stokes scattering is proportional to the mechanical mode's average phonon occupation $\bar{n}$, while Stokes scattering is proportional to $\bar{n}+1$ and hence can become an important contribution at low $\bar{n}$.  When both scattering rates are equal, no further cooling is possible, leaving the mechanical mode in thermal equilibrium with the optical bath.  For optimal detuning of the cooling laser ($\Delta_{\text{opt}}=-\omega_m \sqrt{1+\kappa^2/4 \omega_m^2}$), this temperature limit of the optical bath in units of mechanical quanta is given by $n_{\text{ba}}=(\kappa/4\omega_m)^2$ in the resolved-sideband regime.

As shown in Fig.~\ref{fig:setup}b, the process by which the mechanical motion comes into thermal equilibrium with the optical bath can be observed directly by monitoring the Stokes and anti-Stokes sideband amplitudes.  In our experiment, we collect the light that is Raman-scattered by the cavity directly from the red-detuned cooling laser and perform a heterodyne measurement to separate the sideband amplitudes.  Our experiments operate in a regime where $\kappa \approx \omega_m$.  Here it is possible to be near the mechanical ground state and at the quantum backaction limit simultaneously. Additionally, collecting the Raman-scattered light for thermometry is practical because the Stokes sideband is only partially suppressed by the cavity.  The initial asymmetry of the sidebands is given by the ratio of cavity susceptibility set by the cooling laser detuning $\Delta$.  Hence, in Fig.~\ref{fig:setup}b the anti-Stokes light dominates initially. As the cooling laser power is increased, the coherent cooling rate $\Gamma_{\text{opt}}$ between the mechanical mode and the optical bath is increased. As the mode nears the ground state, the difference between the bosonic factors $\bar{n}$ and $\bar{n}+1$ in the anti-Stokes and Stokes sidebands manifests itself by modifying the sideband asymmetry~\cite{weinstein2014,safavi-naeni2011b,purdy2014,lee2014,meenehan2015,cohen2015}. Here the mode temperature can be determined directly from the scattered light from the cooling laser and hence does not require detailed knowledge of system parameters. The phonon occupation is given by a rate equation that describes the mechanical mode's response to both its environment (at temperature $n_0$) and $n_{\text{ba}}$~\cite{aspelmeyer2014}, which can be written:
\begin{equation}
\label{eq:temp}
\bar{n}(\Gamma_{\text{opt}}) = \frac {n_0 \Gamma_0 + n_{\text{ba}} \Gamma_{\text{opt}}} {\Gamma_0+\Gamma_{\text{opt}}}
\end{equation}
When $\Gamma_{\text{opt}}$ starts to dominate over the mechanical mode's coupling $\Gamma_0$ to its environment, laser cooling begins while maintaining the initial sideband ratio. Once the product $n_{\text{ba}}\Gamma_{\text{opt}}$ exceeds the mechanical decoherence rate $n_0 \Gamma_0$, cooling ceases at the backaction limit ($\bar{n}=n_{\text{ba}}$) where the Stokes and anti-Stokes rates are equal.  Here the competing Stokes and anti-Stokes scattering accomplish nothing but allowing shot noise fluctuations of the cooling laser to set a finite temperature~\cite{marquardt2007,wilson-rae2007}.

Our optomechanical cavity consists of the optical mode of a Fabry-Perot cavity with $\kappa = 2 \pi \times 2.6$ MHz coupled to the motion of the $\omega_m = 2 \pi \times 1.48$ MHz mode of a 500 $\mu$m square by 40 nm thick Si${}_3$N${}_4$ drum resonator (Fig.~\ref{fig:setup}c). The cooling laser is injected into a 10-ppm-transmission mirror. The Raman-scattered light preferentially couples out the second, 100-ppm-transmission mirror, where it is collected via heterodyne detection. An auxiliary locking laser is injected in an orthogonal polarization into the 100 ppm mirror. The cavity is anchored to the base of a dilution refrigerator and light is coupled into the cavity via free space through a narrow cryogenic beam path designed to filter 300 K blackbody radiation~\cite{kuhn2014}. The optical bath is coupled to the mechanical mode at $\Gamma_{\text{opt}} \le 2 \pi \times 30$ kHz, proportional to cooling laser power. The mechanical mode is also coupled to its thermal environment (consisting of cryostat temperature, locking laser RPSN, and other effects), which, in units of mechanical quanta, has a temperature $n_0 = k_B T_0 / \hbar \omega_m \sim 10^3$ corresponding to a temperature $T_0$ (to be determined below), where $k_B$ is Boltzmann's constant and $\hbar$ is the reduced Planck's constant. The high quality-factor Si${}_3$N${}_4$ membrane provides a very low coupling rate to $n_0$, with $\Gamma_0 = 0.18$ Hz, measured via ringdown of a mechanical excitation. 

\begin{figure}
\scalebox{0.9}{\includegraphics{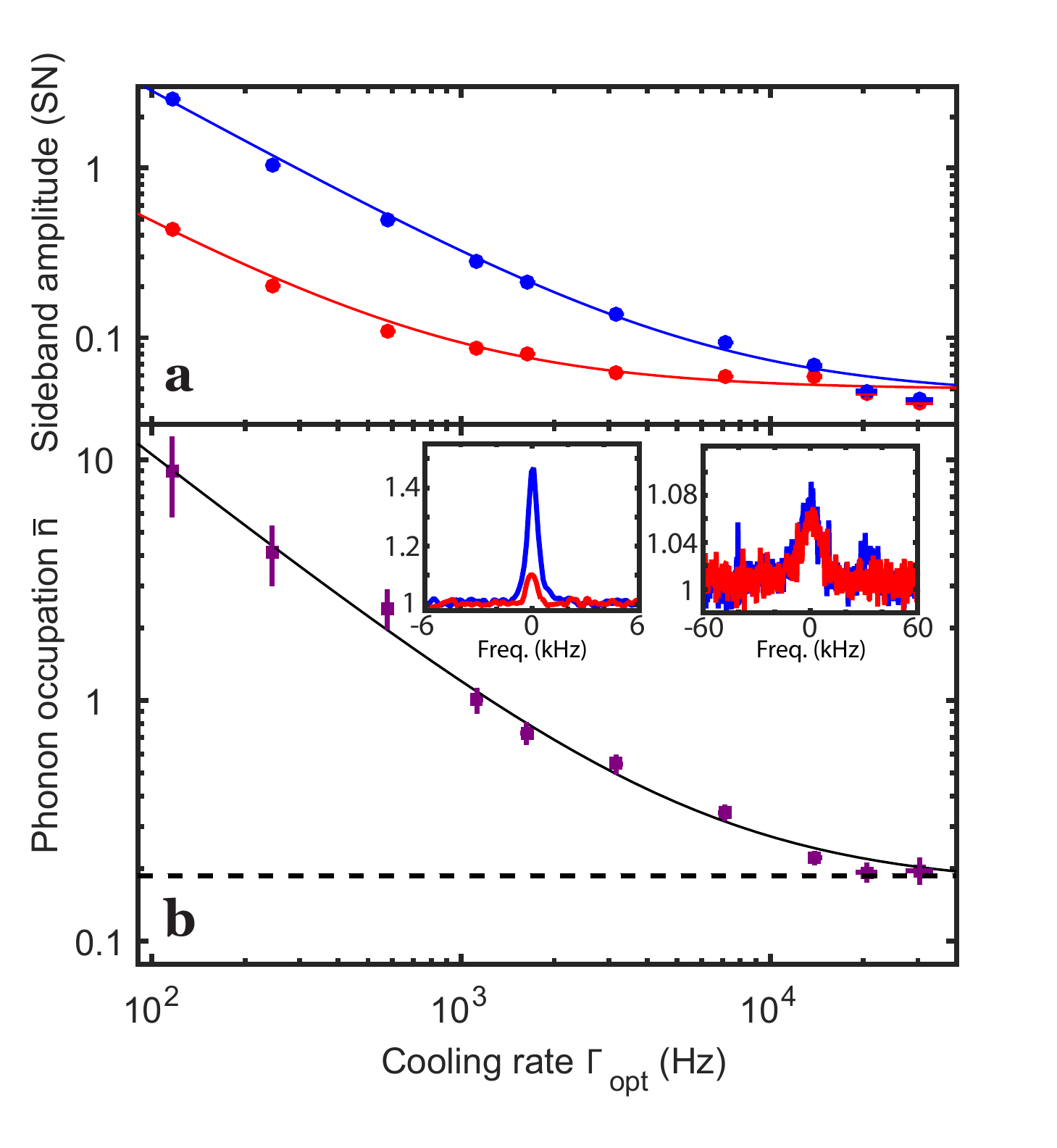}}
\caption{Reaching the quantum backaction limit of sideband cooling near $\Delta_{\text{opt}}$. a) Sideband amplitudes. As $\Gamma_{\text{opt}}$ is increased, the ratio of Stokes (red circles, data; red line, fit) and anti-Stokes (blue circles, data; blue line, fit) sideband amplitudes approaches one. Sideband amplitude is normalized to shot noise (SN). The systematically small values of the sideband amplitudes at largest $\Gamma_{\text{opt}}$ and their small effect on the measurement of $\bar{n}$ are discussed in the main text. b) Minimum temperature. Mechanical occupancy $\bar{n}$ (purple squares) saturates at $\bar{n} = 0.20 \pm 0.02$ for the largest $\Gamma_{\text{opt}}$, in agreement with the quantum backaction limit (dashed black), which is at $n_{\text{ba}}=0.18$ for $\Delta = - 2 \pi \times 1.62$ MHz. A fit to the data (solid black) can be used to infer minimum phonon occupation, as well as the bath temperature $T_0$ of the mode by extrapolation to $\Gamma_{\text{opt}} = 0$. Insets) Overlaid Stokes (red) and anti-Stokes (blue) mechanical spectra, normalized to shot noise. Spectra are third-smallest (left) and third-largest (right) data points.}
\label{fig:cooling}
\end{figure}

Collecting the Raman-scattered light in heterodyne detection allows for thermometry of the mechanical mode without additional probe lasers, since both sidebands are visible in the heterodyne spectrum (Fig.~\ref{fig:cooling}~insets at low and high $\Gamma_{\text{opt}}$). A simultaneous fit to both mechanical sidebands with a common $\omega_m$ (separation from the heterodyne beat note) and $\Gamma_{\text{opt}}$ gives Stokes and anti-Stokes sideband amplitudes normalized to the off-resonant background set by the shot noise of the cooling laser---let the ratio of these amplitudes be $R$. By extrapolating our data to $\Gamma_{\text{opt}} = 0$, we fit the amplitude ratio $s$ due to cavity susceptibility only. Now, we can directly compute $\bar{n}^{-1} = \frac{R}{s}-1$. For each cooling experiment, $\Delta$ is determined by the extrapolated parameter $s$, a function only of cavity susceptibility with otherwise independently measured parameters. A typical cooling experiment is shown in Fig.~\ref{fig:cooling}. As the cooling laser power is increased, $\Gamma_{\text{opt}}$ increases. Monitoring Stokes and anti-Stokes sideband amplitudes (Fig.~\ref{fig:cooling}a and insets) shows the classical regime of constant sideband ratio transitioning to the quantum backaction limit, with equal sideband heights as the mechanical resonator comes into equilibrium with the optical bath. Inferred $\bar{n}$ (Fig.~\ref{fig:cooling}b) shows saturation of the temperature at $n_{\text{ba}}$ (dashed line). For low $\Gamma_{\text{opt}}$, the situation is analogous to previous optomechanical sideband cooling demonstrations. Classically, one expects an inverse relation between $\bar{n}$ and $\Gamma_{\text{opt}}$.  Typically when deviations from this expectation are observed they result from effects such as physical heating due to absorption of light, entering the strong coupling regime, or interference from classical amplitude or phase noise on the cooling laser~\cite{aspelmeyer2014,purdy2012}. However, we find a final temperature in close agreement with $n_{\text{ba}} = 0.18$, determined by independently-measured cavity parameters; the last data point in the cooling curve for Fig.~\ref{fig:cooling} is our lowest measured $\bar{n}=0.20\pm0.02$ for $\Delta = - 2 \pi \times 1.62$ MHz.  

\begin{figure}
\scalebox{1}{\includegraphics{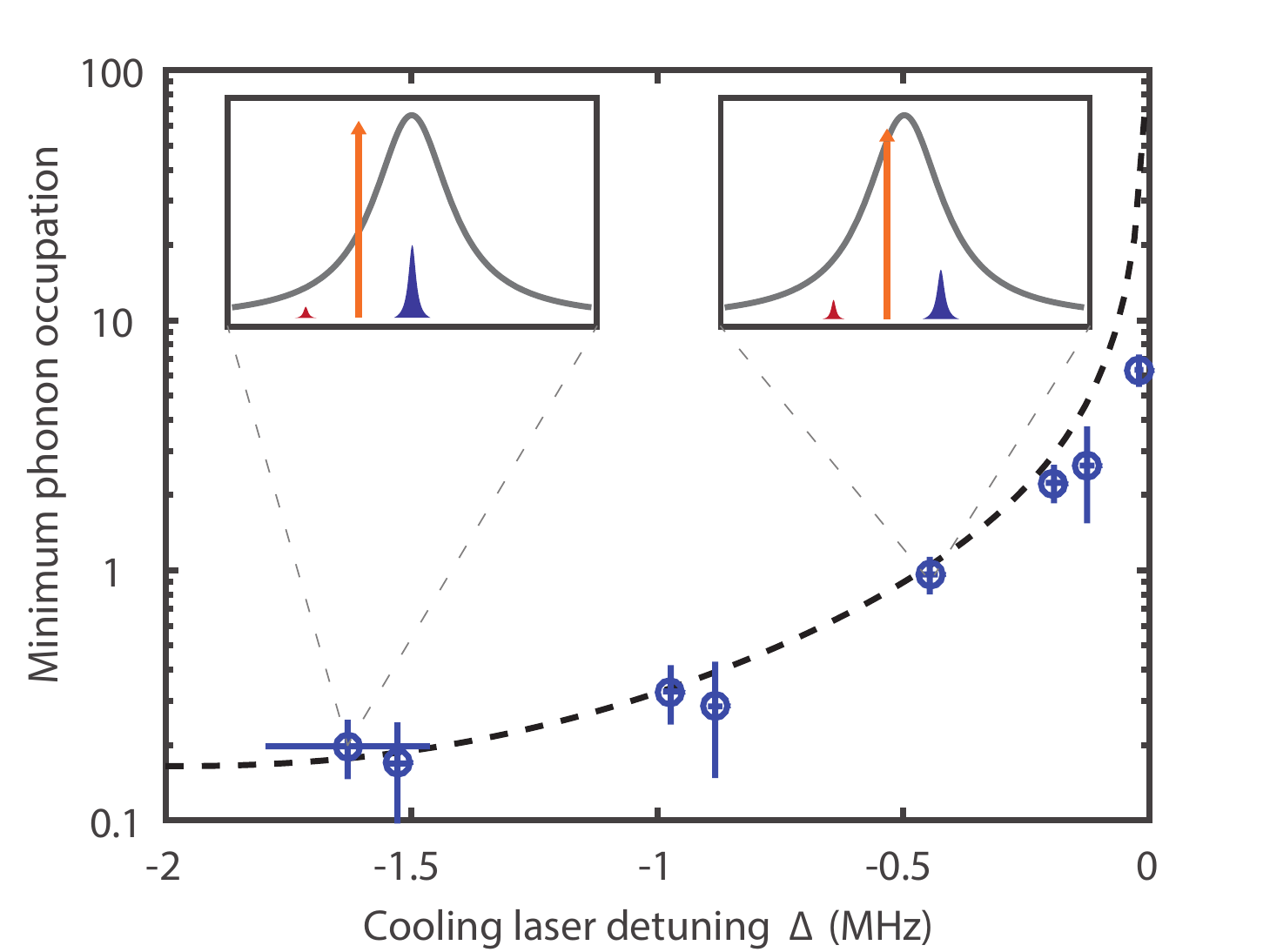}}
\caption{Saturation of the quantum backaction limit. Data (blue open circles) are the fits to the lower limit of $\bar{n}$ in cooling curves such as Fig.~\ref{fig:cooling}. $n_{\text{ba}}$ (dashed line) is given by Eqn.~\ref{eq:limit}.  Insets: Ratio of sidebands in the classical cooling regime for $\Delta \simeq -1.5$ MHz (left) and $\Delta \simeq - 0.5$ MHz (right). A larger ratio in cavity susceptibility allows a lower final $n_{\text{ba}}$.}
\label{fig:detuning}
\end{figure}

For an arbitrary detuning of the cooling laser from the cavity, the quantum backaction limit is expected to change as the sideband resolution is modified.  Specifically, the quantum backaction limit $n_{\text{ba}}$, expressed as an average phonon occupation, takes the form~\cite{marquardt2007}
\begin{equation}
\label{eq:limit}
n_{\text{ba}}(\Delta) = - \frac{(\omega_m+\Delta)^2+(\kappa/2)^2}{4 \omega_m \Delta}
\end{equation}
We have repeated the cooling experiment for a variety of $\Delta$, and can demonstrate saturation of the $\bar{n} = n_{\text{ba}}$ quantum backaction limit at a variety of minimum temperatures (Fig.~\ref{fig:detuning}). The divergence of $n_{\text{ba}}$ as $\Delta \rightarrow 0$ corresponds to the RPSN condition at $\Delta = 0$, where there is no sideband cooling or heating but shot noise in the amplitude quadrature of the light drives the mechanical motion~\cite{purdy2013}.

We must also consider the effect of potential classical noise sources on the measurement.  Overall, the fact that $\bar{n}$ saturates at the expected $n_{\text{ba}}$ is one indicator that classical amplitude and phase noise are not significant systematic errors in the measurement; we have modeled the functional dependences of a variety of forms of classical noise as a function of $\Delta$ and find they will generally make the apparent $\bar{n}$ lie above or below the theoretically predicted $n_{\text{ba}}$~\cite{jayich2012,safavi-naeini2013b,lee2014,purdy2014}.  Nonetheless, we complete a number of independent checks.  First, we independently measure the level of classical amplitude (phase) noise on the cooling laser, finding it to be 0.2\% (2\%) of shot noise at 5 $\mu$W, a representative power at which $\bar{n}=n_{\text{ba}}$.  An analysis of the sideband cooling data that explicitly includes effects from cooling laser amplitude and phase noise changes the final measured phonon occupation in Fig.~\ref{fig:cooling}b by $\Delta \bar{n}^{\text{laser}} \approx 0.006$, less than half the size of the statistical error in the measurement. Another systematic error is the presence of off-resonant substrate mechanical modes that rise above the shot noise floor~\cite{zhao2012, purdy2012, yu2014}.  The normalization of sideband amplitude in our analysis to the off-resonant shot noise level is affected by this noise, causing both sideband amplitudes to lie below the fit in Fig.~\ref{fig:cooling}a. However, since $\bar{n}$ is a function of the sideband ratio, not their absolute amplitudes, it is not strongly affected, and again leads to a small $\Delta \bar{n}^{\text{sub}} \approx 0.006$. Additional confirmation of the substrate noise's small effect is that the mechanical sidebands (Fig.~\ref{fig:cooling}~inset) retain a Lorentzian lineshape~\cite{jayich2012}.  Because both laser noise and substrate noise are small, we otherwise do not include the effects of classical noise directly in the data presentation. 

For each cooling curve, sideband thermometry and knowledge of $\Gamma_0$ allow for extrapolation of $n_0$, and therefore the effective temperature $T_0$ of the mechanical mode.  We find a constant bath temperature of $T_0 = 360$ mK. Beyond the cryostat temperature (70 mK), we can partially trace its origin to RPSN from the locking laser (170 mK), with the balance due to effects whose origin is not completely known. As RPSN from the locking laser was the dominant thermal effect, improved detection of weaker powers would allow for operation at a lower $T_0$. Additionally, increasing locking laser power from 0.6 $\mu$W to 4.8 $\mu$W adds RPSN that raises $T_0$, but does not affect laser cooling to $\bar{n} = n_{\text{ba}}$. Importantly, the lack of power-dependence in $T_0$ suggests that material absorption is not a dominant effect.  This demonstration is key to advancing cryogenic compatibility of these optomechanical devices, which especially for sub-kelvin temperatures can suffer from limited thermalization to cryostat base temperature and heating due to material absorption~\cite{meenehan2014,kuhn2014}.   

The visibility of the mechanical motion against the imprecision noise floor in units of shot noise (Fig.~\ref{fig:cooling}a) is defined by a low total collection efficiency, independently measured to be $\epsilon = 0.04$ based on detection of squeezed light produced by the device. For this work, this excess noise was inconsequential because we only require the relative amplitude of the Stokes and anti-Stokes sidebands.  However, other quantum measurement applications rely on the absolute imprecision of the measurement.  The low efficiency in our measurements was due to heterodyne visibility, cavity losses, propagation loss, and detector efficiency.  The dominant contribution was heterodyne visibility, which can be improved with better alignment.

The work shown here explores the quantum backaction limit of optomechanical sideband cooling, and demonstrates a mechanical resonator in thermal equilibrium with an optical bath rather than its thermal environment. The parameters of our device allow for saturation of the quantum backaction limit while the mechanical mode is in its quantum ground state, coupling mechanical and optical degrees of freedom both in the quantum regime. Recently, similar regimes  of coupling have been reached in demonstrations of squeezed mechanical motion~\cite{wollman2015} and backaction-limited coupling between multiple mechanical modes~\cite{spethmann2015}. We note that the quantum backaction limit can be circumvented by introducing additional couplings to alter the dynamics of the cavity optomechanical system, for example dissipative optomechanical coupling \cite{Elste09,Weiss13}, or measurement and active feedback \cite{Mancini98,Courty01,Genes08}.

\begin{acknowledgments}
This work was supported by DURIP, the DARPA QuASAR program, AFOSR-MURI, AFOSR PECASE, and the National Science Foundation under grant number 1125844. C.R. thanks the Clare Boothe Luce Foundation for support. P.-L.Y. thanks the Taiwan Ministry of Education for support. We thank Antoinne Heidmann and Aurelien Kuhn for design information about free-space optical cryostats.  
\end{acknowledgments}

\bibliography{backactionbib}

\clearpage

\clearpage

\end{document}